\documentclass[twocolumn]{aastex62}

\usepackage{natbib}
\usepackage{hyperref}
\usepackage{floatrow}
\usepackage{amsmath,amssymb}
\usepackage{ulem}
\shorttitle{Kinematics of PDRs in S305}
\shortauthors{N.~K. Bhadari et al.}

\begin{document}

\title{Probing gas kinematics and PDR structure around O-type stars in Sh 2-305 HII region}

\correspondingauthor{N.~K. Bhadari}
\email{Email: naval@prl.res.in}

\author{N.~K. Bhadari}
\affiliation{Physical Research Laboratory, Navrangpura, Ahmedabad 380009, India}
\affiliation{Indian Institute of Technology Gandhinagar Palaj, Gandhinagar 382355, India.}

\author{L.~K. Dewangan}
\affil{Physical Research Laboratory, Navrangpura, Ahmedabad 380009, India}

\author{P.~M. Zemlyanukha}
\affil{Institute of Applied Physics of the Russian Academy of Sciences 46 Ul’yanov str., 603950 Nizhny Novgorod, Russia.}

\author{D.~K. Ojha}
\affil{Department of Astronomy and Astrophysics, Tata Institute of Fundamental Research, Homi Bhabha Road, Mumbai 400005, India.}

\author{I.~I. Zinchenko}
\affil{Institute of Applied Physics of the Russian Academy of Sciences 46 Ul’yanov str., 603950 Nizhny Novgorod, Russia.}

\author{Saurabh Sharma}
\affil{Aryabhatta Research Institute of Observational Sciences, Manora Peak, Nainital 263002, India.}


\begin{abstract}
%
We report an observational study of the Galactic H{\sc ii} region Sh 2-305/S305 using the [C{\sc ii}] 158 $\mu$m line data, which are used to examine the gas dynamics and structure of photodissociation regions.
The integrated [C{\sc ii}] emission map at [39.4, 49.5] km s$^{-1}$ spatially traces two shell-like structures (i.e., inner and outer neutral shells) having a total mass of $\sim$565 M$_\odot$.
The inner neutral shell encompasses an O9.5V star at its centre and has a compact ring-like appearance. 
However, the outer shell is seen with more extended and diffuse [C{\sc ii}] emission, hosting an O8.5V star at its centre and surrounds the inner neutral shell.
The velocity channel maps and position-velocity diagrams confirm the presence of a compact [C{\sc ii}] shell embedded in the diffuse outer shell, and both the shells seem to expand with $v_{\rm exp}\sim$1.3 km s$^{-1}$. The outer shell appears to be older than the inner shell, hinting that these shells are formed sequentially. 
The [C{\sc ii}] profiles are examined toward S305, which are either double-peaked or blue-skewed and have the brighter redshifted component.
The redshifted and blueshifted components spatially trace the inner and outer neutral shell geometry, respectively.
The ionized, neutral, and molecular zones in S305 are seen adjacent to one another around the O-type stars.
The regularly spaced dense molecular and dust clumps (mass $\sim$10--10$^{3}$ M$_{\odot}$) are investigated around the neutral shells, which might have originated due to gravitational instability in the shell of collected materials. 
\end{abstract}



%
\section{Introduction}
\label{sec:intro}
Massive OB-type stars (M$_{\ast}$ $>$ 8 M$_{\odot}$) drive powerful energetics, which allow them to control the physical state of the interstellar medium (ISM). The extreme ultraviolet (EUV; $h\nu\geq$13.6 eV) photons from such massive stars ionize the hydrogen atoms and create H{\sc ii} regions.  
Due to the thermal pressure difference between the ionized region and the surrounding neutral/molecular gas, an H{\sc ii} region continues to grow in size, 
and may trigger star formation in numerous ways \citep[see the review article by][]{Elmegreen98}. 
However, the study of the interaction of massive stars with their surrounding environment is still an open research topic in astrophysics.

The surroundings of the H{\sc ii} regions, where the far ultraviolet (FUV; 6 eV$\leq~h\nu~\leq$13.6 eV) radiation plays a significant role in the heating and/or chemistry of the gas, are referred to as photodissociation regions or photon-dominated regions \citep[PDRs; e.g.,][]{Tielens85a,Tielens85b,Hollenbach99}.
The FUV photons can dissociate the molecules and photoionize those elements having ionization potential less than the Lyman limit.
The gas is primarily neutral in PDRs, but still includes the tracers of singly ionized species (e.g., C$^{+}$, S$^{+}$, Si$^{+}$, and Fe$^{+}$), molecular gas (e.g., H$_{2}$ and CO), and Polycyclic Aromatic Hydrocarbon (PAH) emission \citep[e.g.,][]{Hollenbach99,Kaufman06}.
Hence, the PDRs specify the transition zone between ionized and molecular gas \citep[see Figure~1 in][]{Tielens85a}.
Being a major coolant in PDRs, the $^{2}$P$_{3/2}$--$^{2}$P$_{1/2}$ fine structure line of ionized carbon, [C{\sc ii}] 158 $\mu$m offers a unique probe of kinematic and radiative interaction of massive stars with their surroundings \citep[e.g.,][and references therein]{Abel07,Schneider20}.

In this paper, we employed the observations of 158 $\mu$m [C{\sc ii}] line to study the gas dynamics of PDRs in a promising Galactic H{\sc ii} region, Sh 2-305 (hereafter, S305) powered by two O-type stars. In addition to the study of the spatial-kinematic structure of PDRs in S305, this paper presents the evidence of expanding [C{\sc ii}] shells (also abbreviated as neutral shells in this paper), hinting the applicability of the triggered star formation scenario.

We organize this paper into six sections.
Following the introduction in this section, we present an overview of the Galactic H{\sc ii} region S305 in Section~\ref{sec:overview}.
Section~\ref{sec:obser} presents the description of data-sets used in this work. 
The observational results are presented in Section~\ref{sec:results1}.
The implications of our derived findings are discussed in Section~\ref{sec:discussion}. 
Finally, we summarize our major outcomes in Section~\ref{sec:summary}.
%
%
\section{Overview of the S305 H{\sc ii} region}
\label{sec:overview}
\begin{figure*}
\includegraphics[width=\textwidth]{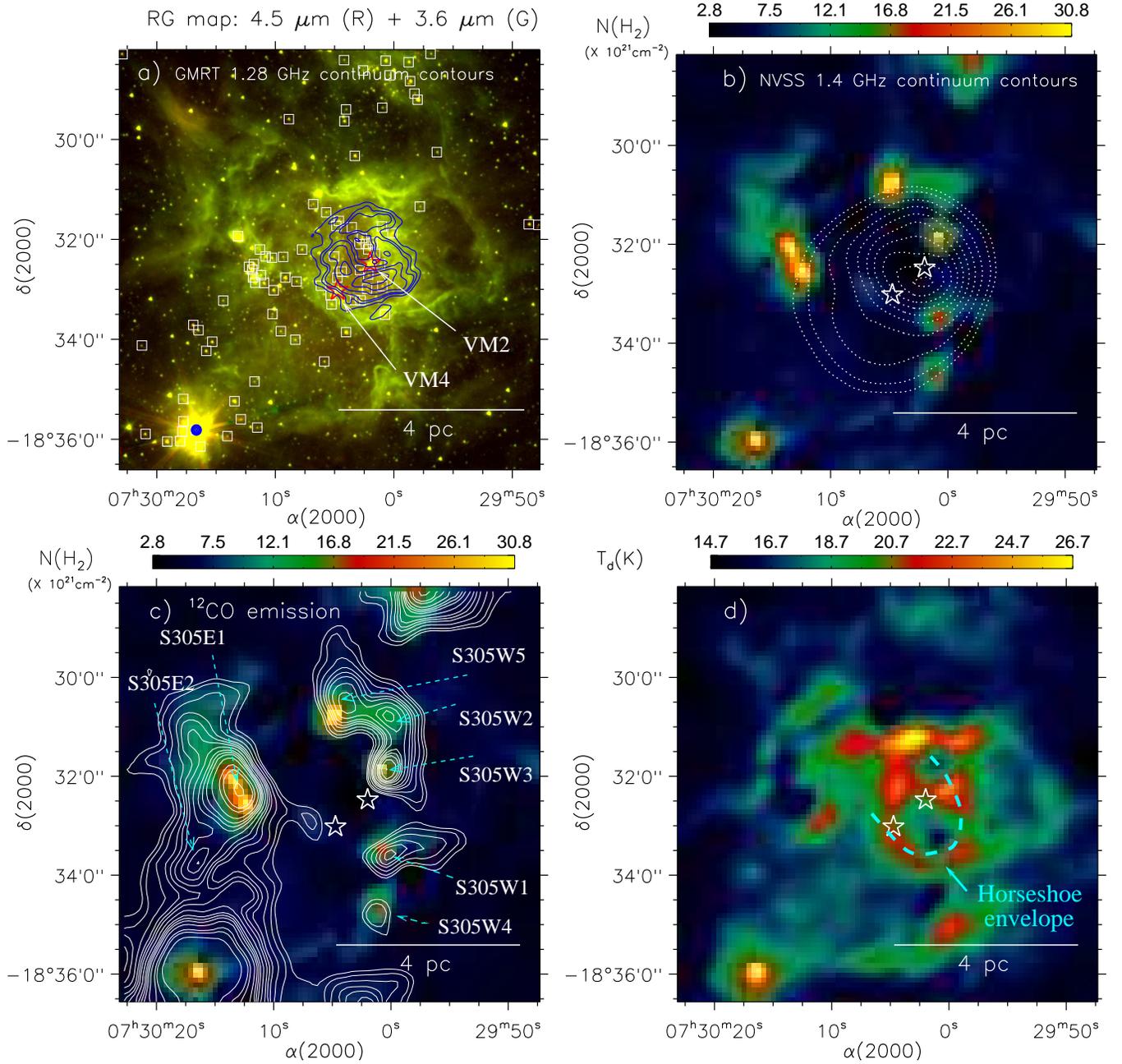}
\caption{(a) Two-color composite image (Red: {\it Spitzer} 4.5 $\mu$m and Green: {\it Spitzer} 3.6 $\mu$m) of the S305 H{\sc ii} region. 
The GMRT 1.28 GHz continuum emission contours (in blue) are overlaid with the 
levels of (0.4, 0.46, 0.5, 0.6, 0.7, 0.8, 0.9, 0.98) $\times$ 22 mJy beam$^{-1}$, where 1$\sigma~\sim$0.74 mJy beam$^{-1}$. 
The positions of YSOs \citep[from][]{Pandey20} are marked by $\square$. 
(b) {\it Herschel} column density map overlaid with the NVSS 1.4 GHz contours. 
The contour levels are (0.15, 0.2, 0.3, 0.4, 0.5, 0.6, 0.7, 0.8, 0.9, 0.98) $\times$ 144 mJy beam$^{-1}$.
(c) Overlay of FUGIN $^{12}$CO(1--0) emission (velocity $\sim$[39, 49] km s$^{-1}$) contours on the {\it Herschel} column density map.
The contour levels are (0.2, 0.22, 0.25, 0.28, 0.3, 0.32, 0.35, 0.38, 0.4, 0.42, 0.45, 0.5, 0.53, 0.6) $\times$ 112.56 K km s$^{-1}$.
The positions of the 850 $\mu$m continuum clumps \citep[from][]{sreenilayam14} are marked by arrows and labeled (see Figure~5 in their paper).
(d) {\it Herschel} temperature map. 
A dashed curve (in cyan) indicates the footprint of the horseshoe envelope \citep[see][for more details]{Dewangan20a}. 
In each panel, the positions of previously known massive O-type stars are marked by star symbols and labeled in panel ``(a)". 
A scale bar referring to 4 pc (at a distance of 3.7 kpc) is also displayed.} 
\label{figx1}
\end{figure*}

%
The S305 H{\sc ii} region is a part of a large molecular cloud complex located at a distance of $\sim$3.7 kpc \citep[][hereafter Paper I]{Pandey20}. 
The molecular cloud associated with S305 has been depicted in a velocity range of [40, 48] km s$^{-1}$ \citep[][hereafter Paper II]{Dewangan20a}. 
In Figure~\ref{figx1}a, we show a two-color composite map ({\it Spitzer} 4.5 $\mu$m (red); {\it Spitzer} 3.6 $\mu$m (green) images) 
of S305, displaying an extended shell-like structure. 
At the centre of the shell-like structure, two massive O-type stars \citep[O8.5V: VM4 and O9.5V: VM2;][]{vogt75,chini84} are located and their positions are shown by star symbols in Figure~\ref{figx1}a.
The positions of Young Stellar Objects (YSOs; from Paper I) and the Giant Metrewave Radio Telescope (GMRT) 1.28 GHz continuum contours (from Paper II) are also overlaid on the color composite image.

In Figure~\ref{figx1}b, we present the {\it Herschel} column density map overlaid with the NRAO VLA Sky Survey \citep[NVSS; resolution $\sim$45$''$;][]{condon98} 1.4 GHz continuum emission contours, allowing us to infer the spatial distribution of dense condensations against the ionized gas.
Figure~\ref{figx1}c displays the overlay of the FUGIN\footnote[1]{FOREST Unbiased Galactic plane Imaging survey with the Nobeyama 45-m telescope}\citep[resolution $\sim$20$''$;][]{umemoto17} $^{12}$CO(1--0) molecular emission contours on the {\it Herschel} column density map. 
The previously known dust clumps \citep[from][]{sreenilayam14} are also labeled in Figure~\ref{figx1}c. 
Noticeable YSOs are also found toward the dust clumps and molecular condensations, which are distributed in a 
regularly spaced manner around the S305 H{\sc ii} region.
The extended structure traced in the {\it Herschel} temperature map of S305 is shown in Figure~\ref{figx1}d. 
Collectively, Figure~\ref{figx1} shows the multi-wavelength view of S305, which was already presented and described in Papers~I and~II.

Previously in Papers~I and~II, the S305 H{\sc ii} region was proposed as a candidate site of triggered star formation. 
It was primarily investigated based on the observed morphology, dynamical age of H{\sc ii} region, and the fragmentation time scale of the associated molecular shell. However, a direct observational proof of the triggered star formation in S305 remains ambiguous. 

%
\section{Data sets}
\label{sec:obser}
%

\subsection{SOFIA [C{\sc ii}] observations}
In this paper, we used the science ready 158 $\mu$m [C{\sc ii}] line data cube of S305 (Project ID: 06\_0226; PI: Loren Dean Anderson), which was obtained from the InfraRed Science Archive\footnote[2]{https://irsa.ipac.caltech.edu/applications/sofia}. 
The line observations were taken with the Stratospheric Observatory for Infrared Astronomy (SOFIA)/upGREAT\footnote[3]{upGREAT\citep{Risacher16} is an enhanced version of the German Receiver for Astronomy at Terahertz Frequencies (GREAT).} instrument. 
The [C{\sc ii}] line data have a half-power beamwidth of 14\rlap.{$''$}1 and a velocity resolution of 0.385 km s$^{-1}$ \citep[see][for data reduction procedure]{Anderson19,Schneider20}. We smoothed the original line data cube with a Gaussian function having 3 pixels half-power beamwidth (i.e., 21\rlap.{$''$}15, where 1 pixel corresponds to 7\rlap.{$''$}05), which improved the image sensitivity. The resulting resolution of the line data cube is 25\rlap.{$''$}4.
\subsection{Ancillary Data}
We used multi-scale and multi-wavelength data sets, which were obtained from different existing surveys (e.g., 
the Warm-{\it Spitzer} GLIMPSE360 Survey \citep[$\lambda$ = 3.6 and 4.5 $\mu$m; resolution $\sim$2$''$;][]{benjamin03,whitney11},
Wide-field Infrared Survey Explorer \citep[{\it WISE}; $\lambda$ = 12 $\mu$m; resolution $\sim$6\rlap.{$''$}5;][]{Wright10},
the NVSS, and the FUGIN survey ($^{12}$CO/$^{13}$CO (J=1--0); resolution $\sim$20$''$--21$''$)). 
Apart from these data sets, we also utilized the {\it Herschel} temperature and column density maps (resolution $\sim$12$''$) from \citet{molinari10} which were generated for the {\it EU-funded ViaLactea project}. 
The Bayesian {\it PPMAP} method \citep{marsh15,marsh17} was applied to build these {\it Herschel} maps. 
The GMRT radio continuum map at 1280 MHz (resolution $\sim$10$''$) was taken from Paper II.
%
\section{Results}
\label{sec:results1}
In Paper~II, molecular condensations, PAH emission, dust clumps, and H$_{2}$ emission have been traced toward the horseshoe envelope surrounding the ionized shell, where noticeable YSOs are also found (see Figure~\ref{figx1}). The ionized shell is traced by the GMRT 1.28 GHz continuum emission (see Figure~\ref{figx1}a), while the footprint of the horseshoe envelope is shown by a dashed curve in Figure~\ref{figx1}d. 
In the following sections, we study new observations of the 158 $\mu$m [C{\sc ii}] line toward S305, enabling us to examine the spatial and velocity structure of PDRs in our target site.
\subsection{The spatial-kinematic structure of PDRs in S305}
\label{sec:results}
\begin{figure*}
\includegraphics[width=\textwidth]{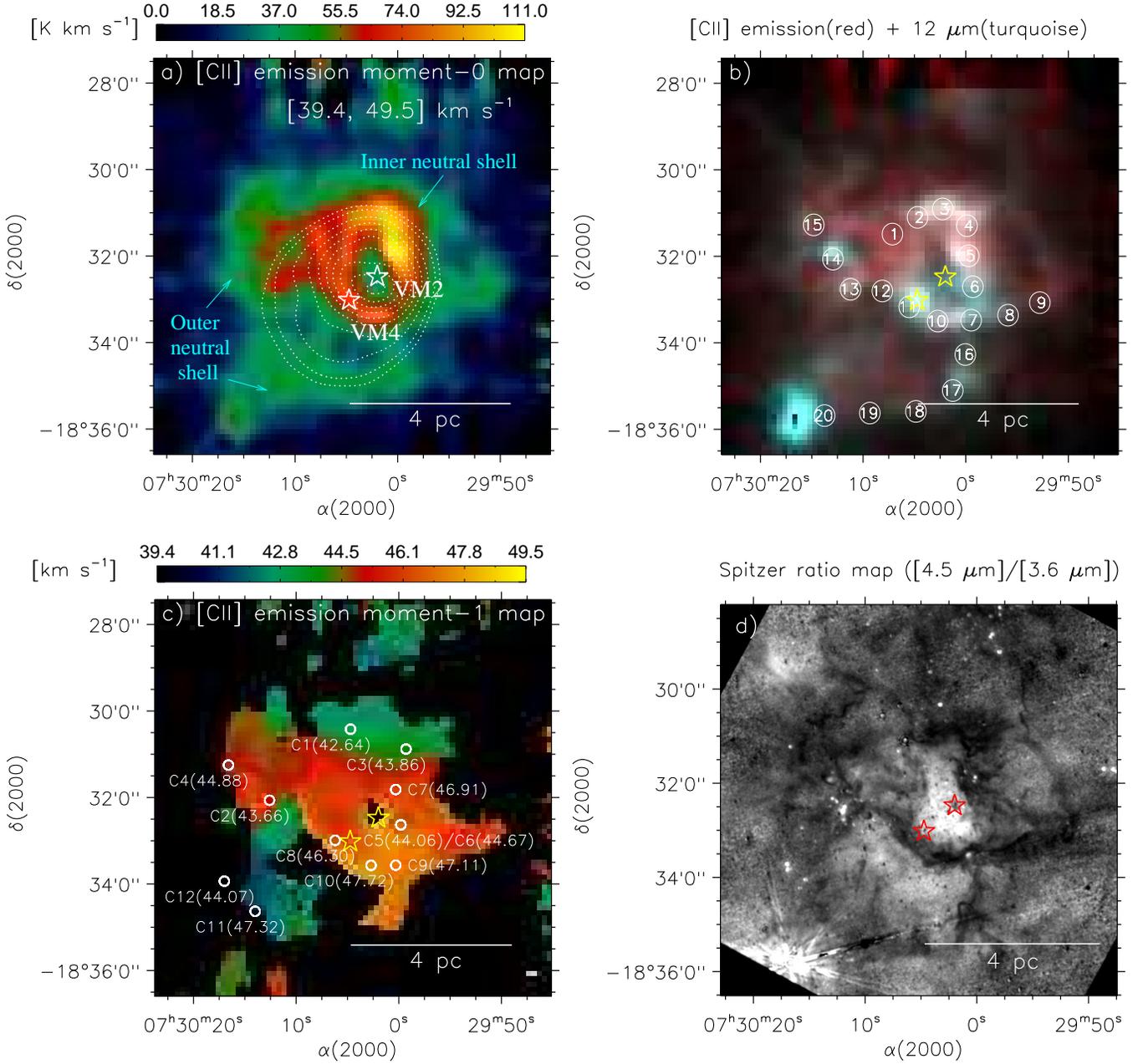}
\caption{(a) The panel displays the [C{\sc ii}] integrated intensity (moment-0) map of S305. 
The map is overlaid with the NVSS 1.4 GHz continuum emission contours, which are the same as in Figure~\ref{figx1}b. 
(b) Two-color composite image (Red: [C{\sc ii}] moment-0 map and Turquoise: {\it WISE} 12 $\mu$m image) of the S305 H{\sc ii} region. 
The twenty circles (radius =15$\arcsec$ each) represent the areas for which the gas spectra are extracted (see Figure~\ref{figx4}).
(c) [C{\sc ii}] intensity weighted mean velocity (moment-1) map overlaid with the positions of $^{12}$CO(2--1) molecular clumps \citep[from][]{azimlu11}. The clumps are highlighted by open circles and corresponding velocities are also labeled.
(d) The {\it Spitzer} ratio map of 4.5 $\mu$m/3.6 $\mu$m emission \citep[reproduced from][]{Dewangan20a}.
In all the panels, the stars are the same as in Figure~\ref{figx1}.}
\label{figx2}
\end{figure*}

In Figure~\ref{figx2}a, we show an integrated intensity (moment-0) map of the [C{\sc ii}] emission around S305, where the [C{\sc ii}] emission is integrated over a velocity range of [39.4, 49.5] km s$^{-1}$. 
The NVSS 1.4 GHz continuum emission contours are also overlaid on the [C{\sc ii}] emission map.
The integrated [C{\sc ii}] emission map hints the presence of two shell-like structures (i.e., inner and outer neutral shells) in S305.
These shells can be regarded as the PDRs in S305 region (see Section~\ref{sec:intro}).
The inner neutral shell, having a compact ring-like appearance, is evident with a high intensity value of [56, 110] K km s$^{-1}$, and extends toward the eastern direction. 
A depreciation of the [C{\sc ii}] emission (i.e., a cavity), the radio continuum peak emission, and 
the position of the massive O9.5V star (i.e., VM2) spatially coincide, and are found at the centre of the inner neutral shell. 
The outer neutral shell is traced with the diffuse and more extended [C{\sc ii}] emission having a lower intensity value of [20, 55] K km s$^{-1}$, and surrounds the inner neutral shell. 
The massive O8.5V star (i.e., VM4) appears to be located at the centre of the outer neutral shell. 
This particular morphology is also seen in the H$\alpha$ image from Paper II (see Figure~9 therein), which suggests that VM4 may be the major source of feedback in the dusty H{\sc ii} region S305.	
The other [C{\sc ii}] emission peaks are also seen toward the north and south-east directions and lie toward the locations of the previously identified clusters of YSOs (e.g., Paper I).

We show a two-color composite map (Red: [C{\sc ii}] moment-0 map; Turquoise: {\it WISE} 12 $\mu$m image) of S305 in Figure~\ref{figx2}b.
The spatially extended structures seem to match well in both images. 
In Figure~\ref{figx2}b, we have also marked twenty circular regions, where the gas spectra are extracted and examined.
Figure~\ref{figx2}c presents the [C{\sc ii}] moment-1 map that allows us to study the spatial distribution of the mean velocity of emitting gas. A nearly linear velocity gradient can be seen in the north-south direction. However, one can notice that the inner neutral shell is redshifted compared to the blueshifted outer/extended diffuse shell. The [C{\sc ii}] moment-1 map is also overlaid with the positions of previously identified $^{12}$CO(2--1) molecular clumps and their velocities \citep[][see also Paper II]{azimlu11}. 
The velocities of these clumps are in agreement with the velocities of the [C{\sc ii}] gas at corresponding positions. 
Nearly eight molecular clumps are found to be spatially coincident with the inner neutral shell. 
Six out of eight molecular clumps (i.e., c5--c10) are seen in the direction of the horseshoe envelope (see Figure~\ref{figx1}d). 

In order to compare the spatial structure traced by the [C{\sc ii}] emission and the infrared continuum images, we present the {\it Spitzer} ratio map of 4.5 $\mu$m emission to 3.6 $\mu$m emission in Figure~\ref{figx2}d. 
A detailed discussion on the ratio map is given in Paper~II.
The bright and dark emission zones in the ratio map detect the ionized emission and the PDR walls, respectively.
One can notice from Figures~\ref{figx2}a and \ref{figx2}d, that the boundaries of the [C{\sc ii}] shells are well traced by the dark structures in the ratio map.
Thus, the analysis of multiwavelength data confirms the presence of two shells.

Figure~\ref{figx3} presents the velocity channel contours of the [C{\sc ii}] emission overlaid on the {\it WISE} 12 $\mu$m image, enabling us to explore the gas motion. We find the spatial match of the [C{\sc ii}] emission with the infrared structure traced in the {\it WISE} 12 $\mu$m image. 
In general, the {\it WISE} 12 $\mu$m image is known to cover the prominent PAH features at 11.3 $\mu$m.
In the channel maps, based on the visual inspection, we marked the boundaries of two [C{\sc ii}] shells by dotted circles.
The larger dotted circle (centre coordinates; RA = 07$^{\rm h}$ 30$^{\rm m}$ 04$^{\rm s}$.51, Dec = -18$\degr$ 32$\arcmin$ 32$\arcsec$.72, and radius = 205$\arcsec$) encloses the [C{\sc ii}] emission of the outer diffuse shell. The outer [C{\sc ii}] shell is clearly seen in the velocity range from 42.1 to 44.1 km s$^{-1}$, and has an incomplete circular (or broken ring) appearance.
On the other hand, the smaller circle (centre coordinates; RA = 07$^{\rm h}$ 30$^{\rm m}$ 02$^{\rm s}$.39, Dec = -18$\degr$ 32$\arcmin$ 19$\arcsec$.95, and radius = 105$\arcsec$) surrounds the inner [C{\sc ii}] shell, which is evident in the velocity channel maps from 45.2 to 47.5 km s$^{-1}$. 
In the direction of inner shell structure, the [C{\sc ii}] channel maps show that the [C{\sc ii}] emission is blueshifted in the northern direction, while it is strongly redshifted in the southern direction. 
This implies that the gas is moving toward (away from) the observer in the northern (southern) directions \citep[e.g.,][]{Mookerjea21}.
Overall, these results suggest the expansion of the gas in PDRs (see Section~\ref{sec:discussion} for detailed discussion).

In order to compare the kinematics of different gas components, we have analysed the spectral profiles of $^{12}$CO/$^{13}$CO(1--0) and [C{\sc ii}] gas in Figure~\ref{figx4}.
Figure~\ref{figx4} displays the line profiles of the [C{\sc ii}], $^{12}$CO(1--0), and $^{13}$CO(1--0) emission toward twenty circular regions (radius =15$\arcsec$ each) in the direction of S305. These regions are primarily selected toward the direction of prominent infrared features and the [C{\sc ii}] emission (see circles in Figure~\ref{figx2}b). 
Each spectrum is produced by averaging the emission over the circular regions. 
In the direction of the compact ring feature (i.e., positions 1--15), [C{\sc ii}] line profiles are brighter and are either blue-skewed or have double-peaked structures compared to the $^{12}$CO(1--0), and $^{13}$CO(1--0) profiles.
The [C{\sc ii}] line profiles show double peaks at $\sim$43 and $\sim$47 km s$^{-1}$, and a dip around $\sim$44 km s$^{-1}$. 
On the other hand, the $^{12}$CO(1--0) and $^{13}$CO(1--0) profiles are single-peaked and lie between the double peaks of the [C{\sc ii}] spectra.
Interestingly, all the double-peaked profiles have a brighter redshifted component than the blueshifted one (see Section~\ref{sec:discussion} for more discussion).

In Figure~\ref{figx5}, we have shown the [C{\sc ii}] line profiles (similar to those presented in Figure~\ref{figx4}) with color coded redshifted and blueshifted components. From the velocity channel maps (Figure~\ref{figx3}), we have identified the velocity ranges of these components. The blueshifted component ([39.4, 44.4] km s$^{-1}$) is found to be associated with the outer shell structure, while the redshifted component ([44.4, 49.5] km s$^{-1}$) traces the inner shell structure. 
One can notice from Figure~\ref{figx5} that the peak of the redshifted component (i.e., 45.6 km s$^{-1}$; see panels 2, 3, and 4) shifts toward higher velocities as we move southwards from the northern direction (see Figure~\ref{figx2}b). However, the blueshifted component peaks at 43.3 km s$^{-1}$ in all the positions, and does not show much variation along the velocity axis. These results suggest the noticable gas expansion in the inner shell compared to the outer shell. However, considering the simplistic case, we assume a similar expansion velocity for both the shells to estimate their expansion timescales (see Section~\ref{sec:masses}).

Based on the selected velocity ranges of redshifted and blueshifted components from Figure~\ref{figx5}, we have examined the moment maps.
Figures~\ref{figx6}a and \ref{figx6}b show the moment-0 and moment-1 maps for the redshifted component, respectively. The moment-0 map shows the inner shell structure having the compact ring like appearance, while the moment-1 map in Figure~\ref{figx6}b reveals the velocity gradient along the north-south direction of the inner shell.
The velocity at the shell's rim is bluer than the centre, which favors the shell expansion.
Generally, the outward displacement of the shell with increasing or decreasing velocity is referred to as the kinematic signature of an expanding shell \citep[e.g.,][see Figure~6 therein]{Pabst19}.
Similarly, Figures~\ref{figx6}c and \ref{figx6}d display the moment-0 and moment-1 maps for the blueshifted component, respectively. The outer shell is evident in the moment-0 map (Figure~\ref{figx6}c) and has a broken ring morphology. The moment-1 map in Figure~\ref{figx6}d traces the velocity field in the outer shell. A nearly linear velocity gradient is seen in the east-west direction, which may infer the direction of gas expansion in the outer shell. However, the velocity structure is more complex in the outer shell than the inner shell, which makes it more difficult to interpret.

In order to compare the spatial distribution of different components of gas (i.e., ionized, neutral, and molecular) in S305, we show a three-color (Red: GMRT 1.28 GHz, Green: FUGIN $^{12}$CO(1--0), Blue: SOFIA [C{\sc ii}] 158 $\mu$m) composite image in Figure~\ref{figx7}a.
All the three components of gas are clearly distinguishable. 
In the direction of the observed [C{\sc ii}] emission, the molecular CO is less abundant or mostly absent (see also Figure~\ref{figx4}). 
The distribution of the molecular gas is clumpy rather than a diffuse appearance. 
Figure~\ref{figx7}b displays the [C{\sc ii}] emission contours for the inner and outer shell structures (see Figures~\ref{figx6}a and \ref{figx6}c).
The detailed interpretation of the shell structures is discussed in  Section~\ref{sec:doublepdr}.

To study the spatial-kinematic structures of the PDRs in S305, we have studied the position-velocity (pv) diagrams of different gas components.
Figures~\ref{figx8}a and \ref{figx8}b show the pv diagrams of [C{\sc ii}], $^{12}$CO(1--0), and $^{13}$CO(1--0) emission in two different directions; i.e., perpendicular and along to the line joining the stars VM2 and VM4, respectively. 
The positive offsets in the positions are measured along the south-west and north-west directions for the two pv diagrams.
We have overlaid the emission contours of $^{13}$CO(1--0) (in red) and [C{\sc ii}] (in violet) on the $^{12}$CO(1--0) pv emission map, allowing us to study the spatial-kinematic structures of different components of gas.
The 1.28 GHz radio brightness profile is also shown in each panel, specifying the position of the H{\sc ii} region. 
The detailed outcomes from this analysis are discussed in Section~\ref{sec:discussion}.

\subsection{Physical parameters of the [C{\sc ii}] shells}
\label{sec:masses}

We have estimated the neutral hydrogen column density ($N(\mathrm H)$) and the mass of the [C{\sc ii}] emitting gas toward the compact as well as the extended feature (i.e., inner and outer neutral shells). 
The column density depends on the luminosity ratio of the [C{\sc ii}] line in the PDR to that of the H{\sc ii} region 
as follows \citep[see equation A4 in][]{Kaufman06}
 
\begin{equation}
N(\mathrm H)= \left(\frac{L_{\rm C{\sc II}}(\rm PDR)}{L_{\rm C{\sc II}}(\rm H{\sc II})}\right) \frac{Zn_{e}^{1/3}\Phi_{49}^{1/3}f_{\rm C^{+}}}{15} \times 10^{21}~({\rm cm^{-2}}),
\end{equation}

where $\Phi_{i}=10^{49}\Phi_{49}$ s$^{-1}$ is the absolute luminosity of ionizing EUV photons, $f_{\rm C^{+}}$ is the fraction of singly ionized carbon, $Z=1$ relates to the standard abundance or metalicity, and $n_{e}$ is the number density of electrons. 
From the [C{\sc ii}] integrated intensity map, we have estimated the value of $\frac{L_{\rm C{\sc II}}(\rm PDR)}{L_{\rm C{\sc II}}(\rm H{\sc II})}$ around 9.
In this exercise, we consider that the [C{\sc ii}] emission, which spatially overlaps with the 1.28 GHz continuum emission (i.e., H{\sc ii} region), belongs to the H{\sc ii} region and the rest of emission corresponds to the part of PDRs.

For the typical values of $Z=1$, $n_{e}=500~{\rm cm^{-3}}$, $\Phi_{i}=10^{49}$ s$^{-1}$, and $f_{\rm C^{+}}=0.1$ 
\citep[adopted from][]{Kaufman06,Kirsanova20},
the derived $N(\mathrm H)$ and the total mass of the [C{\sc ii}] emitting gas in the PDRs are $\sim$$5\times10^{20}~{\rm cm^{-2}}$ and $\sim$565 M$_{\odot}$, respectively. The mass calculation is based on the assumption that PDRs have the shape of layers on the sphere, and all the material is in the form of  neutral hydrogen.
We calculated the layer widths from the pv diagrams (Figure~\ref{figx8}). 
To compute the total mass of the shells, we used the mass-column density relation of $M_{\rm shell}= m_{\rm H}~a_{\rm shell}~\Sigma N(\mathrm H)$, where  $a_{\rm shell}$ is the shell area, $m_{\rm H}$ is the mass of hydrogen atom, and $\Sigma N(\mathrm H)$ is the integrated column density over the shell area.
In this exercise, the uncertainty caused by the noise is around 10\%, but the assumptions of different parameters \citep[e.g., $n_{e}$, $f_{\rm C^{+}}$, and $\Phi_{i}$ from][]{Kaufman06,Kirsanova20} can change the outcomes accordingly.
The hydrogen column density (or shell mass) can also be derived from the [C{\sc ii}] emission, which first requires the estimation of [C{\sc ii}] column density ($N$([C{\sc ii}])).
Since the optically thin [$^{13}$C{\sc ii}] line is not detected in present observations, we can assume [C{\sc ii}] emission as optically thin and then compute $N$([C{\sc ii}]) using equation 4 in \citet{Kirsanova20}. Assuming a typical excitation temperature of $T_{ex}$=120 K, the observed average $N$([C{\sc ii}]) is $\sim$$4\times10^{17}~{\rm cm^{-2}}$.
Now, using the gas-phase carbon abundance of [C/H]= $1.6\times10^{-4}$ \citep{Sofia04}, we evaluated the $N(\mathrm H)$ as $\sim$$2\times10^{21}~{\rm cm^{-2}}$, which is almost close to the previously calculated $N(\mathrm H)$ value of $\sim$$5\times10^{20}~{\rm cm^{-2}}$.

We have also estimated the expansion timescale of each [C{\sc ii}] shell, considering that the expansion is caused by a continuous flow of stellar winds \citep[e.g.,][]{Castor75,Weaver77}.
The radii ($R_{\rm s}$) of the inner and outer shells are computed from the pv diagrams and the velocity channel maps (see dotted circles in Figure~\ref{figx3}), and are $\sim$1.9 pc and $\sim$3.7 pc, respectively.
We have evaluated the expansion velocity of $v_{\rm exp}\sim$1.3 km s$^{-1}$, assuming that the shells expand in both directions \citep[e.g.,][see Section~3.2.2 and Appendix B therein]{Pabst20}. 
Here, we also note that the expansion velocity can range from 1--4 km s$^{-1}$ (see pv diagrams in Figure~\ref{figx8}), and for the calculation purpose, we have used our calculated value of $\sim$1.3 km s$^{-1}$.
Considering a similar expansion velocity of $v_{\rm exp}\sim$1.3 km s$^{-1}$ for both the shells, the expansion timescales of the inner and outer shells are determined to be $\sim$0.9 Myr and $\sim$1.7 Myr, respectively.
To compute the expansion timescale, we used the following equation \citep[e.g.,][]{Weaver77, Pabst20} 
\begin{equation}
t_{\rm exp}\simeq 0.6 \left(\frac{ R_{\rm s}}{\rm 1~pc}\right) \left(\frac{\rm 1~km~s^{-1}}{v_{\rm exp}}\right) {\rm Myr}.
\end{equation}

Here, one can note that the systematic uncertainties in the estimation of shell extent, the expansion velocity, and the expansion timescale are about 30--50\%.

%
%
%
%
\section{Discussion}
\label{sec:discussion}
%
In this paper, for the first time, we examined the structure and gas dynamics of PDRs in S305 using the neutral gas tracer, 158 $\mu$m [C{\sc ii}] line.
The neutral gas is traced in the velocity range of [39.4, 49.5] km s$^{-1}$, which is consistent in velocity with the molecular gas (see Figure~\ref{figx4} and Paper II). The ionized, neutral, and molecular gas are spatially distributed in the hierarchy manner (i.e., adjacent to one another; see Figure~\ref{figx7}a) in S305, which agrees well with the one-dimensional models of PDR \citep[e.g.,][see Figure~1 therein]{Tielens85a}. 
In Figure~\ref{figx7}a, the boundary, traced by the [C{\sc ii}] gas, is well distinct to that of ionized gas and molecular CO. This is probably possible because the CO molecules, in the direction of massive stars and our line of sight are mostly dissociated and the carbon is ionized, which contribute to the [C{\sc ii}] emission in PDRs around S305. 
It seems that the [C{\sc ii}] emission traces the unshielded H$_{2}$ gas in PDRs, where molecular gas tracer CO is photodissociated.

\subsection{Double [C{\sc ii}] shells in PDRs around S305}
\label{sec:doublepdr}
\begin{figure*}
\includegraphics[width=\textwidth]{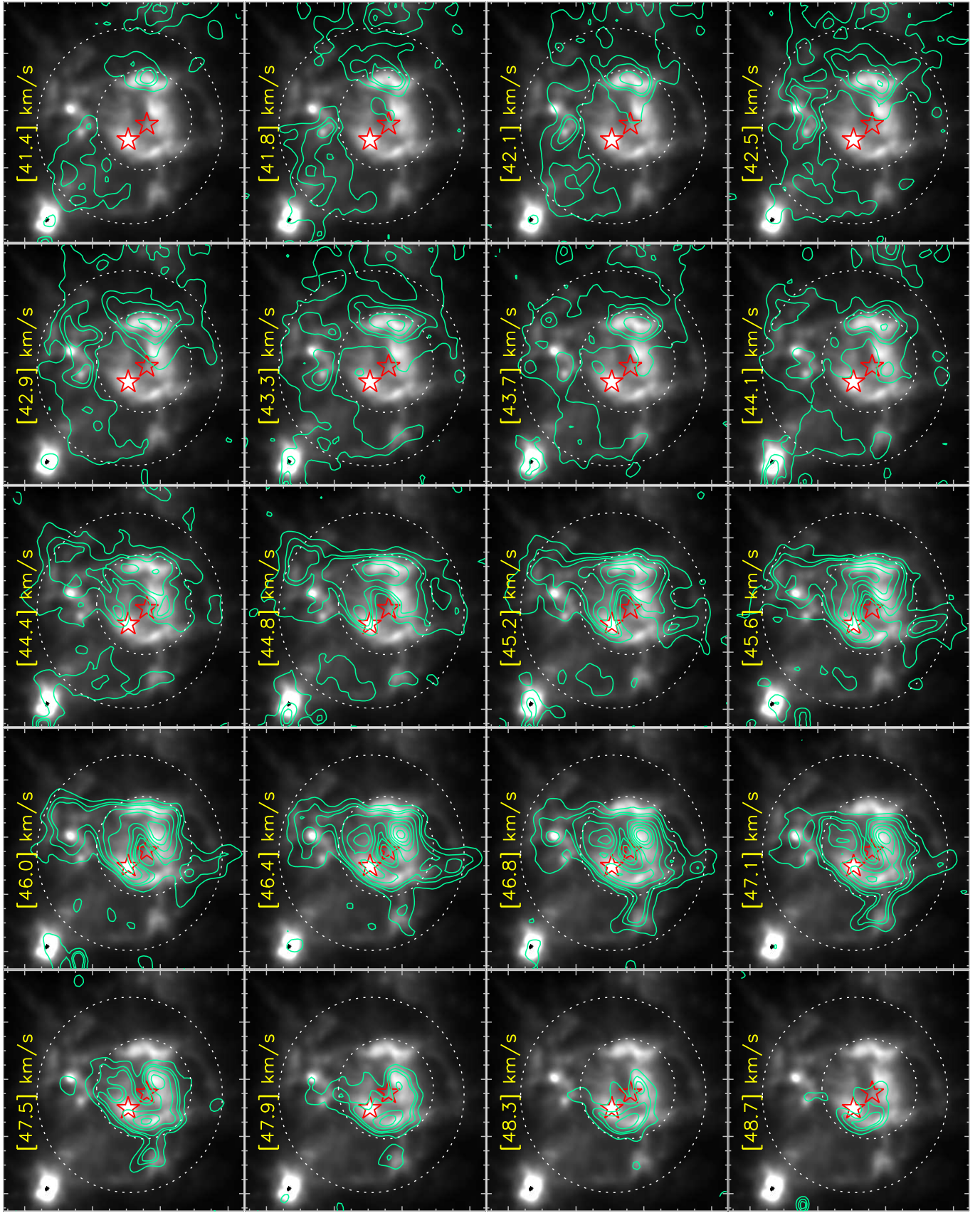}
\caption{Velocity channel contours of the [C{\sc ii}] emission overlaid on the {\it WISE} 12 $\mu$m image. 
The velocity value is indicated in each panel (in km s$^{-1}$). 
The contours (in spring green) are shown with the levels of 5, 8, 10, 15, 20, 25, 30, and 35 K.
The dotted circles represent the footprint of inner and outer [C{\sc ii}] shells.
In each panel, the positions of known massive stars are marked by star symbols.} 
\label{figx3}
\end{figure*}

\begin{figure*}
\includegraphics[width=1\textwidth]{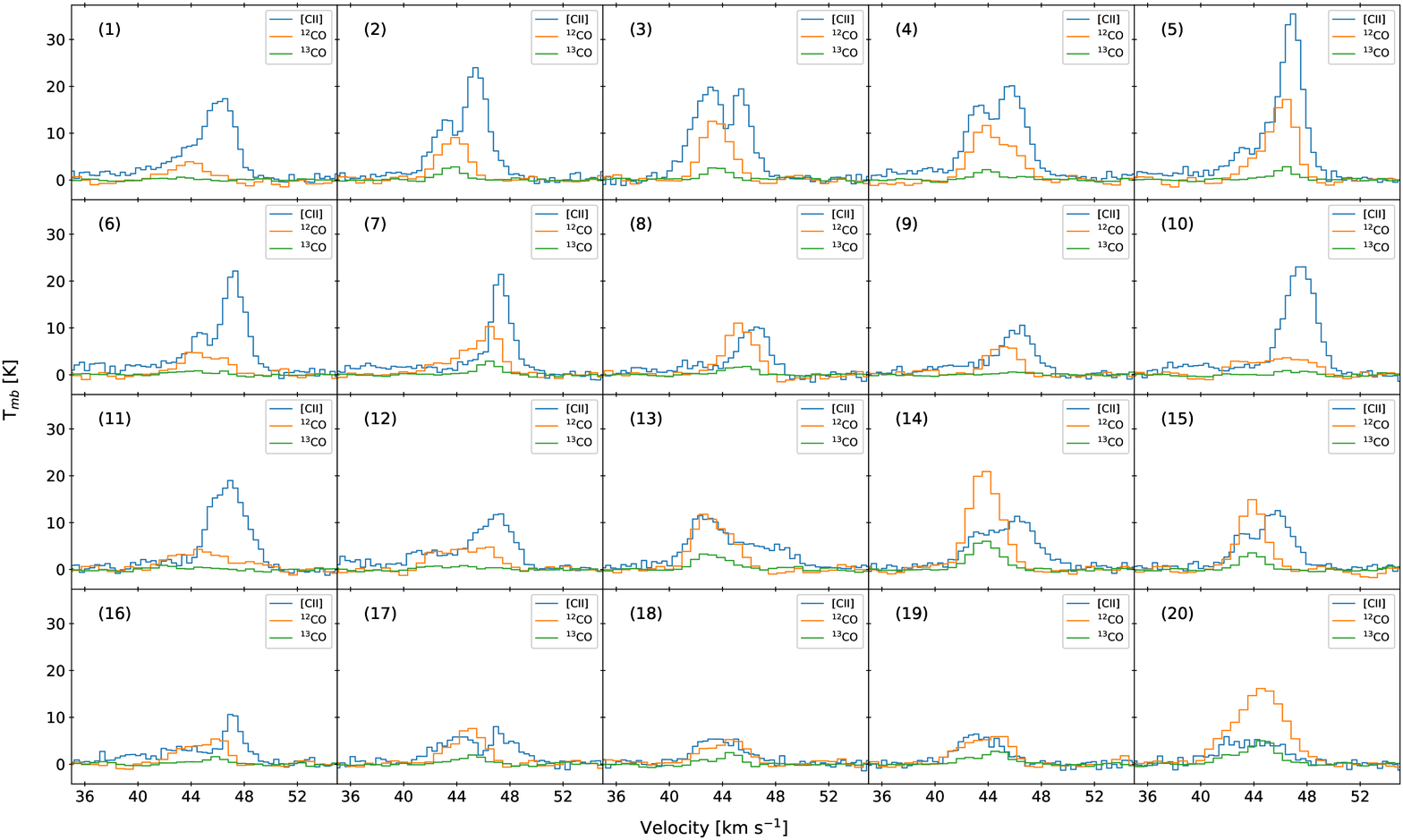}
\caption{Profiles of the [C{\sc ii}], $^{12}$CO(1--0), and $^{13}$CO(1--0) emission toward twenty circular regions (radius =15$\arcsec$ each) in S305 (see circles in Figure~\ref{figx2}b). The corresponding circle number is marked in each panel.}
\label{figx4}
\end{figure*}

\begin{figure*}
\includegraphics[width=1\textwidth]{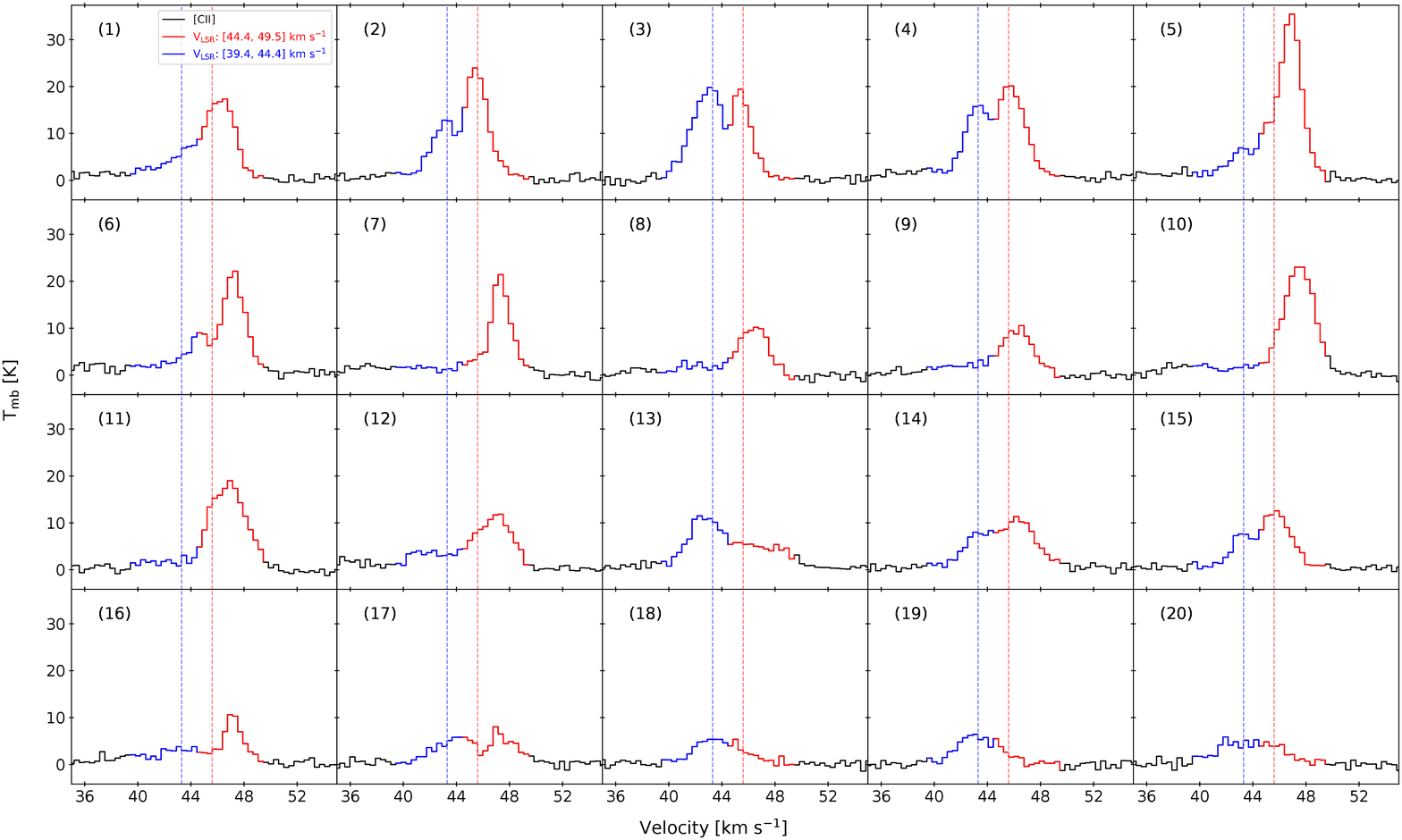}
\caption{Profiles of the [C{\sc ii}] emission, similar to those shown in Figure~\ref{figx4}. The blue and red components in the [C{\sc ii}] profiles correspond to the velocity range of [39.4, 44.4] km s$^{-1}$ and [44.4, 49.5] km s$^{-1}$, respectively. These velocity ranges are used to disentangle the inner and outer [C{\sc ii}] shells (see Section~\ref{sec:results} for more details).
The vertical dashed lines in blue and red represent the velocities of 43.3 and 45.6 km s$^{-1}$, respectively.
}
\label{figx5}
\end{figure*}

\begin{figure*}
\includegraphics[width=1\textwidth]{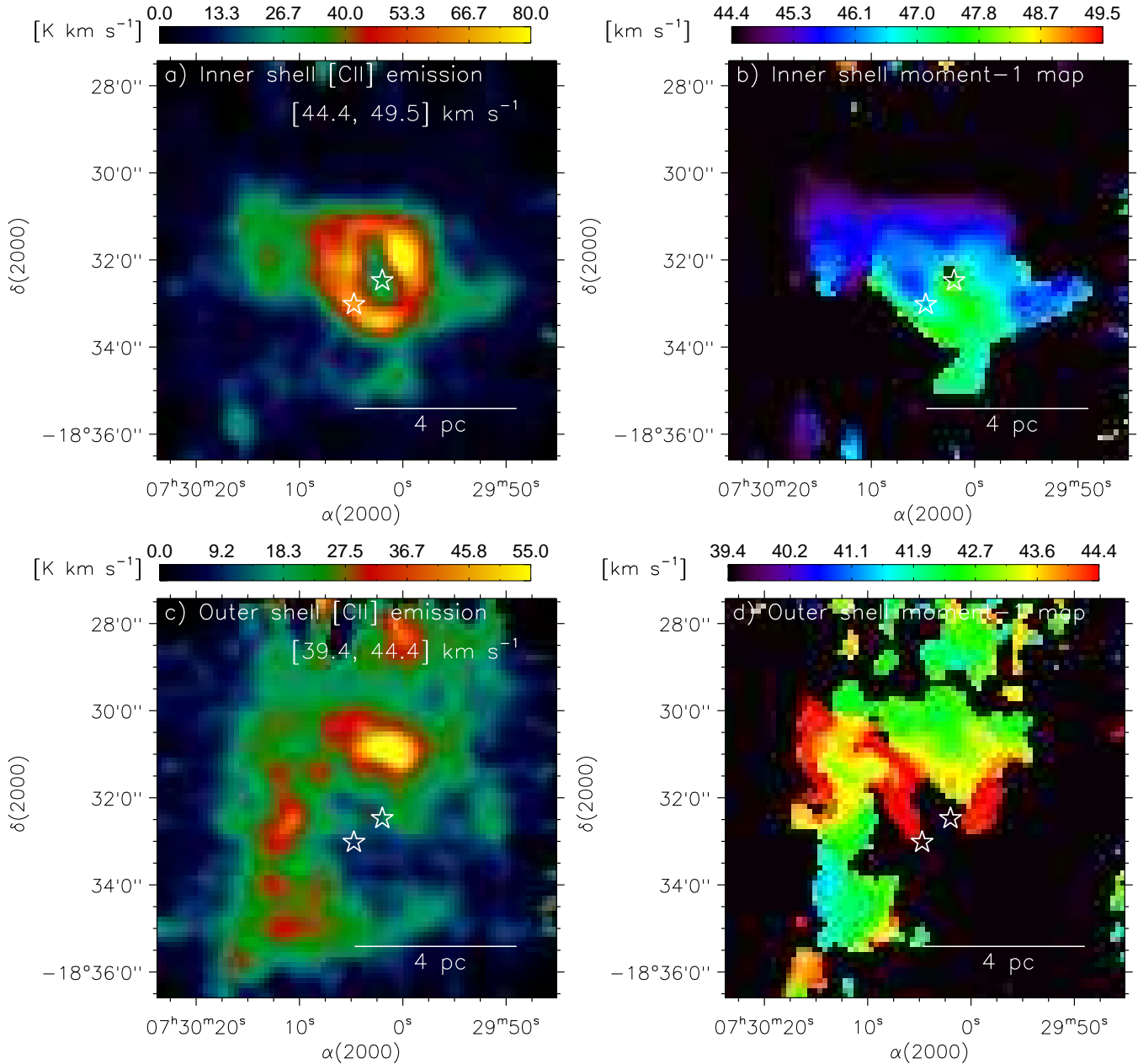}
\caption{Top panels display (a) moment-0 map and (b) moment-1 map of the [C{\sc ii}] emission in the velocity range of [44.4, 49.5] km s$^{-1}$ (see red component of the [C{\sc ii}] profiles in Figure~\ref{figx5}).
Bottom panels show the similar maps as of top panels, but for the outer [C{\sc ii}] shell (see blue component of the [C{\sc ii}] profiles in Figure~\ref{figx5}).
In all the panels, the stars are the same as in Figure~\ref{figx1}.
}
\label{figx6}
\end{figure*}

\begin{figure*}
\includegraphics[width=11.5cm]{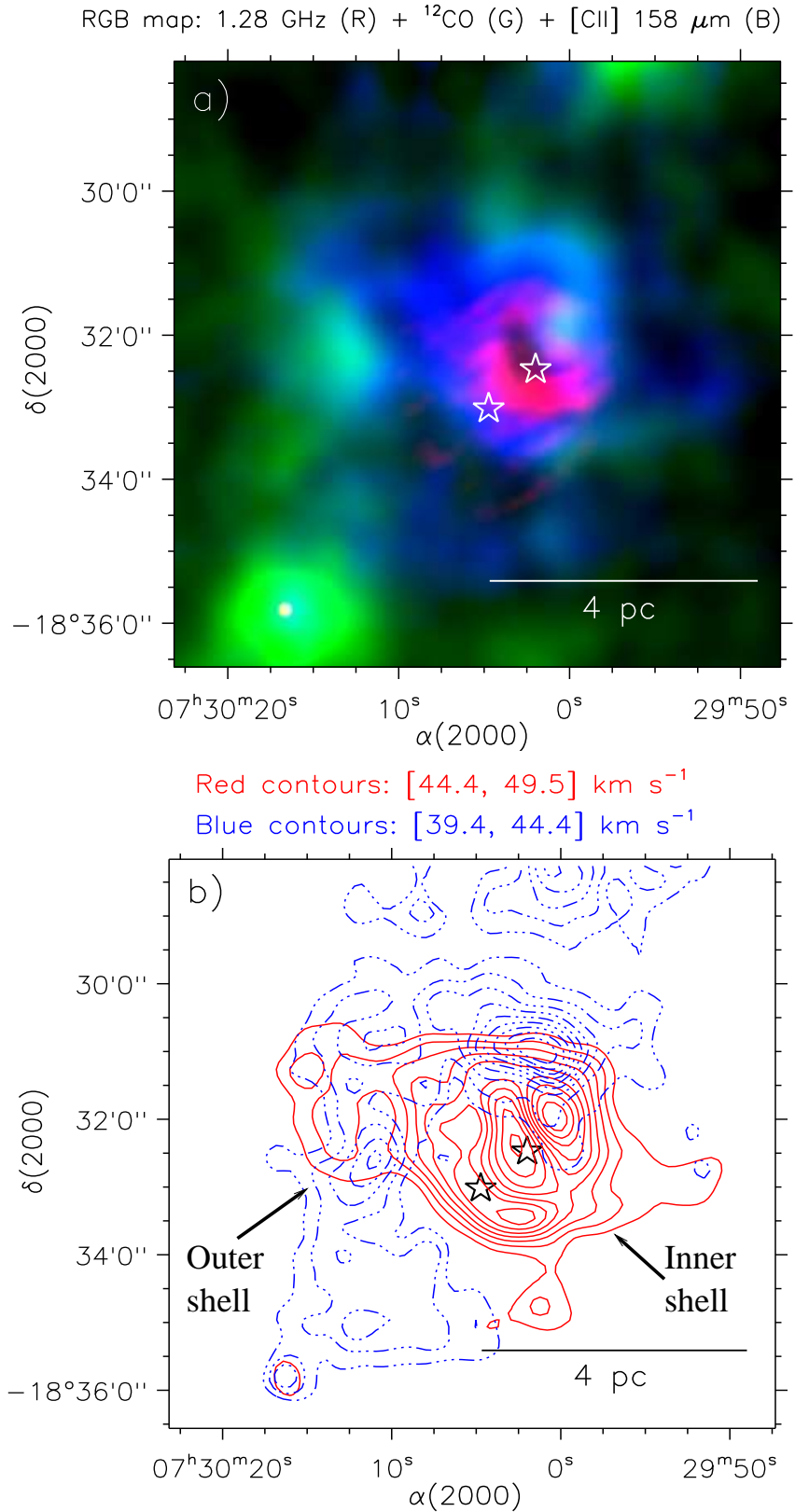}
\caption{(a) Three-color (Red: GMRT 1.28 GHz, Green: FUGIN $^{12}$CO(1--0), Blue: SOFIA [C{\sc ii}] 158 $\mu$m) composite map of S305. (b) The [C{\sc ii}] emission contours for the inner (in red) and outer (in blue) shell structures.
The red contour levels range from 23.34 to 93.36 K km s$^{-1}$ in steps of 7.78 K km s$^{-1}$, while the blue contour levels range from 19.57 to 65.24 K km s$^{-1}$ in steps of 5.07 K km s$^{-1}$.
The stars are the same as in Figure~\ref{figx1}.
}
\label{figx7}
\end{figure*}

\begin{figure*}
\includegraphics[width=11.5cm]{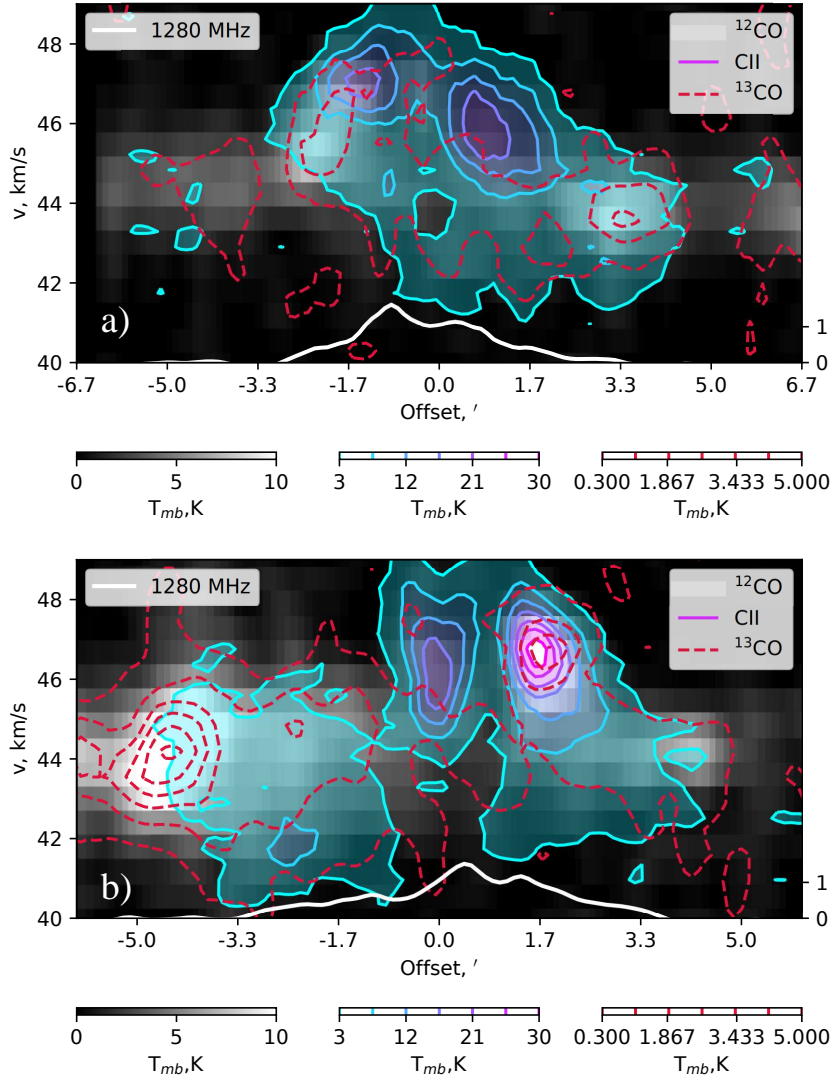}
\caption{Position-velocity diagrams of the $^{12}$CO(1--0) (grey emission), $^{13}$CO(1--0) (dashed contours), and [C{\sc ii}] (thick contours) emission; (a) perpendicular to the line joining VM2 and VM4, (b) along the line joining VM2 and VM4. 
The corresponding color bars are shown below the panels. 
In each panel, the 1280 MHz radio continuum emission profile (a thick white curve) is also shown, which is multiplied by a factor of 100. The radio continuum brightness scale bar is shown in right to each panel and has a unit of mJy beam$^{-1}$. 
}
\label{figx8}
\end{figure*}

The two [C{\sc ii}] shells in S305 are evident from the channel maps shown in Figure~\ref{figx3}.
The outer shell has a more extended diffuse structure and unveils its broken morphology, while the inner shell reveals a compact ring like feature (see Section~\ref{sec:results} and Figure~\ref{figx6}).
Recently, \citet{Kirsanova20} performed the spherically symmetric chemo-dynamical model of expanding H{\sc ii} regions in the site S235.
The pv diagrams shown in Figure~\ref{figx8} are close to the numerical model by \citet{Kirsanova20}. 
In the direction of the line joining stars VM2 and VM4, the shape of the pv diagram is somewhat symmetrical for the individual shells and has a typical double-peaked structure (i.e., similar to the dumbbell shape; see Figure~\ref{figx8}b). 
The footprint of the diffuse shell can be seen in the positive and negative offsets ([C{\sc ii}] contour of 3 K), while the compact shell is seen toward the positive offset ([C{\sc ii}] contours of 6--30 K) only. 
The velocity of the extended diffuse shell is around $\sim$44 km s$^{-1}$ compared to the compact shell having a velocity of $\sim$46.5 km s$^{-1}$.
The molecular gas walls are formed around the diffuse shell as seen in $^{12}$CO(1--0) and $^{13}$CO(1--0) peaks at $\sim$44 km s$^{-1}$. 

The pv diagram (Figure~\ref{figx8}b) also infers the geometric distribution of the two [C{\sc ii}] shells.
In the direction of the compact shell, the molecular gas wall, which is kinematically related to the diffuse shell wall, shows a peak emission at $\sim$46 km s$^{-1}$. This molecular peak emission exactly matches with one of the two [C{\sc ii}] peaks (located at 1$'$.7 offset and  $\sim$46 km s$^{-1}$).
The other [C{\sc ii}] peak related to the compact shell (located at 0$'$ offset and  $\sim$46 km s$^{-1}$) in the pv diagram is situated in the center of the diffuse shell. 
The absence of the molecular wall in this peak suggests that the compact zone is embedded into the extended diffuse shell. 
This particular observation provides the only evidence of geometrically inner (compact shell) and outer (diffuse shell) morphology of the [C{\sc ii}] emitting gas. 
The pv structures in the different directions are far from symmetry, as shown in the pv diagram in the perpendicular direction of the line joining stars VM2 and VM4 (see Figure~\ref{figx8}a). Such asymmetry can arise from the inhomogeneity of the surrounding gas. These kinematical structures may be explained by the triggered formation scenario of the compact zone, where the parental gas is compressed by the expansion of the diffuse shell. 
The H{\sc ii} zone lies at the centre of the diffuse shell as shown by the 1.28 GHz continuum emission profile and is the outcome of EUV flux from both the stars VM2 and VM4 (Paper II).
The gas is being pushed to the observer leading to the doppler-shift of the [C{\sc ii}] emission. The compact H{\sc ii} zone expands to the inner regions of the diffuse shell, which can explain the asymmetry in the pv diagrams. The gas expansion in the inner [C{\sc ii}] shell is clearly evident from the velocity channel maps (Figure~\ref{figx3}) and moment-1 map (Figure~\ref{figx6}b).
However, it is not very significant in case of the outer shell, which although show the south-east to south-west velocity gradient (see Figure~\ref{figx6}d).

The shells are not kinematically homogeneous but are close to the expanding bubbles.
In Section~\ref{sec:masses}, the expansion timescales of the inner and outer shells are determined to be $\sim$0.9 Myr and $\sim$1.7 Myr, respectively. 
The expansion timescales of the shells agree with the dynamical age of the S305 H{\sc ii} region (i.e., $\sim$0.5--1.7 Myr; Paper II), 
and with the average age of the stellar population in S305 (i.e., $\sim$1.8 Myr; Paper I).
These results infer that the inner shell is younger than the outer shell, suggesting that the shells have formed sequentially. This may be possible that the star VM4 played a major role in the formation of the outer shell, while the inner shell formation seems to be linked with the star VM2 (see also Figures~\ref{figx2} and \ref{figx3}). 
Furthermore, based on the morphology and kinematics of the [C{\sc ii}] shells, we cannot rule out the possibility of the influence of the star VM4 in the formation of the star VM2.
However, it requires the proper age estimation of these stars, which is beyond the scope of this paper. 

In the direction of S305, we find the double-peaked line profiles of the [C{\sc ii}] emission compared to the single-peaked line profiles of $^{12}$CO(1--0) and $^{13}$CO(1--0) (see Figure~\ref{figx4}). 
From the analysis of the [C{\sc ii}] spectra taken at different positions in S305 (Figure~\ref{figx5}), we found that the blueshifted and redshifted components spatially trace the outer and inner shell morphology, respectively (see Figure~\ref{figx6}).
It is quite interesting to note that the outer shell which is more diffuse compared to the inner shell, lies in front of the inner shell.
We note that the double-peaked [C{\sc ii}] profiles are dominant in the northern high intensity zone of the inner neutral shell (see Figures~\ref{figx2}b and ~\ref{figx5}), which spatially lies to the overlapping zones of the inner and outer [C{\sc ii}] shells (see also Figure~\ref{figx7}b).
It has been shown in the previous studies that the presence of an enhanced redshifted peak in the significant optically thick lines is the signature of expansion scenario \citep[e.g.,][]{Pavlyuchenkov08}. 
In S305, the majority of the [C{\sc ii}] profiles (Figure~\ref{figx4}) are either double-peaked (with brighter redshifted component) or blue-skewed, suggesting the gas expansion in PDRs, which is further confirmed by the [C{\sc ii}] channel maps and the pv diagrams.

\subsection{Star formation scenario in S305}
\label{sec:triggerstarformation}

In the literature, it has been a common practice to use the mid-infrared images as a tool for qualitative identification of bubble structures or triggered star formation sites \citep[e.g.,][]{Deharveng05,Deharveng10,Zavagno06,Zavagno07}. Different scenarios of triggered/sequential star formation are available 
in the literature \citep[e.g.,][]{Elmegreen98}. The expansion of an H{\sc ii} region may initiate the instability in pre-existing dense regions of gas \citep[i.e., ``radiation driven implosion" scenario;][]{Bertoldi89, Lefloch94} or may accumulate the layers of dust and gas between the ionization and shock fronts which later become gravitationally unstable and form next-generation stars \citep[i.e., ``collect and collapse" scenario;][]{Elmegreen77,Whitworth94}.
Despite of the enumerable wealth of literature in theoretical works, the direct observational evidence of ``triggered/sequential star formation" is still lacking \citep[e.g.,][and references therein]{Deharveng10,Dale15}. 

In the mid-infrared images, the 24 $\mu$m emission correlates to the ionized zone, while the 8 $\mu$m emission traces the boundary of dust condensation and the ionization front.
The shock front can be traced by H$_{2}$ emission. 
In S305, the H$_{2}$ emission is prominently found in the direction of the horseshoe envelope (see Figure~3 in Paper II). 
The presence of regularly spaced massive fragments in the periphery or along the direction of PDRs is evident in the dust emission at 850 $\mu$m (see Figure~\ref{figx1}c), and in the form of molecular clumps (see Figure~\ref{figx2}c). 
These clumps are spatially seen in the direction and/or the boundary of the [C{\sc ii}] shells, which are also kinematically related to the [C{\sc ii}] emitting gas.
The masses of dust and molecular clumps are in the range of M$_{\rm clumps}^{\rm 850~\mu m}$ $\sim$10--10$^{3}$ M$_{\odot}$ \citep[see Table~6 in][]{sreenilayam14} and M$_{\rm clumps}^{\rm CO(2-1)}$ $\sim$10$^{2}$--10$^{3}$ M$_{\odot}$ \citep[see Table~3 in][]{azimlu11}, respectively.
Apart from this, Paper~I identified the massive dense clumps/fragments using the {\it Herschel} column density map. These clumps have size extensions in parsec (i.e., radii $\sim$0.28--1.32 pc) and their masses range from 35 to 1565 M$_{\odot}$. 
Hence, it is possible that during the expansion of an H{\sc ii} region, the dense materials have been collected between the ionization and shock fronts, which now appear in the form of dense fragments. 

Therefore, all the outcomes together confirm the gas expansion in PDRs followed by the collection of materials (e.g., dust and gas) around it. 
The presence of massive neutral shells, dust and gas clumps around H{\sc ii} region S305 strengthens the applicability of ``collect and collapse" scenario as suggested in Paper II.

%
%
\section{Summary and Conclusion}
\label{sec:summary}
%
The present paper is benefited with analysis of the unpublished 158 $\mu$m [C{\sc ii}] line data, which enable us to study the kinematics of gas in PDRs around the H{\sc ii} region S305 hosting two massive O-type stars (i.e., VM2 and VM4).
Using the [C{\sc ii}] line data, the neutral gas in PDRs is examined in a velocity range of [39.4, 49.5] km s$^{-1}$.
Two shell-like structures (i.e., inner: an intense ring-like shell and outer: the extended and diffuse neutral shell) 
are found in the [C{\sc ii}] moment-0 map, and velocity channel maps. 
In the direction of S305, the [C{\sc ii}] line profiles are double-peaked (with enhanced redshifted component), which peaks at $\sim$43 and $\sim$47 km s$^{-1}$ and shows a dip at $\sim$44 km s$^{-1}$. The dip in double-peaked [C{\sc ii}] profiles lie to the peaks of $^{12}$CO/$^{13}$CO(1--0) line profiles. The double-peaked [C{\sc ii}] spectra spatially trace two shells in PDRs around S305.
The pv diagrams confirm that the compact shell is geometrically embedded in the outer diffuse shell, and both the shells are enclosed by the molecular gas walls. 
Based on the [C{\sc ii}] emission, the estimated average neutral hydrogen column density and the mass of the [C{\sc ii}] shells are $\sim$$5\times10^{20}~{\rm cm^{-2}}$ and $\sim$565 M$_{\odot}$, respectively.
The pv diagrams, including the spectral profiles and velocity channel maps, unveil the signatures of gas expansion in PDRs. The expansion velocity of $v_{\rm exp}\sim$1.3 km s$^{-1}$ implies the expansion timescales of $\sim$0.9 Myr and $\sim$1.7 Myr for the inner and outer shells, respectively. This suggests that the two [C{\sc ii}] shells are formed sequentially in S305. 
The consistent velocity and spatial structures of regularly spaced massive dust and molecular fragments to that of the neutral gas/PDRs confirm that these fragments are the outcome of the gravitational collapse of a shell of collected materials.

\section[]{Acknowledgments}
We thank the anonymous reviewer for valuable comments which have improved the scientific quality of the paper.
The research work at Physical Research Laboratory is funded by the Department of Space, Government of India. 
DKO acknowledges the support of the Department of Atomic Energy, Government of India, under Project Identification No. RTI 4002. 
IIZ and PMZ acknowledge the support of the Russian Science Foundation (grant No. 17-12-01256). 
This work is based on observations made with the NASA/DLR Stratospheric Observatory for Infrared Astronomy (SOFIA). SOFIA is jointly operated by the Universities Space Research Association, Inc. (USRA), under NASA contract NAS2-97001, and the Deutsches SOFIA Institut (DSI) under DLR contract 50 OK 0901 to the University of Stuttgart.
This work is based [in part] on observations made with the {\it Spitzer} Space Telescope, which is operated by the Jet Propulsion Laboratory, California Institute of Technology under a contract with NASA. 
This publication makes use of data from FUGIN, FOREST Unbiased Galactic plane Imaging survey with the Nobeyama 45-m telescope, a legacy project in the Nobeyama 45-m radio telescope. 

%
%

%

\end{document}